\documentclass[multphys,vecphys]{svmult}

\usepackage{makeidx}         
\usepackage{graphicx}        
\usepackage{multicol}        
\usepackage[bottom]{footmisc}
\usepackage{amsmath,amssymb,url,color}
\definecolor{red}{rgb}{1,0,0}
\definecolor{green}{rgb}{0,1,0}
\definecolor{blue}{rgb}{0,0,1}

\makeindex             

\begin{document}

\title*{The network approach:\\ 
        basic concepts and algorithms}
\titlerunning{The network approach} 

\author{Pedro G.~Lind}
\authorrunning{P.G.~Lind}
\institute{ICP, Universit\"at Stuttgart, Pfaffenwaldring 27, 
           D-70569 Stuttgart, Germany
           \texttt{lind@icp.uni-stuttgart.de}}

\maketitle



\section{The network approach as a physical problem}

As economists are the experts on consumer attitudes and sociologists 
on human social interaction, physicists are the experts simplifying 
complex problems.
However, one usually does not ask a physicist about stock-market 
forecasts~\cite{stockmarket}, neither does one immediately think of 
a physicist when the issue concerns controlling civilian 
crowds~\cite{helbing}. 
Scientists dealing with many complex real problems do not take
kindly to the propensity of physicists entering their field
of study and proposing overly simplified theories.
One even often hears jokes about physicists assuming chickens to 
be spherical. Though, such kind of simplifications and assumptions
do turn out to be rather helpful in solving a specific real-life 
problem at times.

Of course, not all assumptions can be taken.
They must retain the essential features to explain what we observe. 
Take a balloon filled with air, for instance. The complicated 
system we call `air' comprehends a huge number of molecules of nitrogen, 
oxygen, argon and carbon dioxide, among many other. 
Each one of these molecules is an arrangement of atoms with different 
weights, and even each atom is a system of quarks and electrons 
interacting according to specific forces. 
To explain why a balloon expands 
while being filled with air, you do not need to think about all these 
details.
One just {\it assumes} that air is a set of small spheres traveling in
all directions and colliding among them and with the walls of the 
balloon. 
These collisions are {\it essentially} what is responsible for what we 
call gas pressure. Everything else can be ignored.

This way of dealing with reality, trying to simplify it down to its
elementary components and interactions for explaining a certain
phenomenon is what leads physicists to uncover fundamental laws 
of nature.
And schematically, it is also precisely what the so-called 
{\it network approach} is about: the components of the system under study 
are reduced to elements that retain the essential features we want to 
address and the interactions between such components are represented by 
links joining the elements~\cite{livro2}.
Elements could be molecules with links representing the collisions 
among them, but they could also be seen as persons linked by their
friendship acquaintances or as enterprises connected among them according
to the trades they establish.
\begin{figure}[t]
\begin{center}
\includegraphics*[width=10.0cm,angle=0]{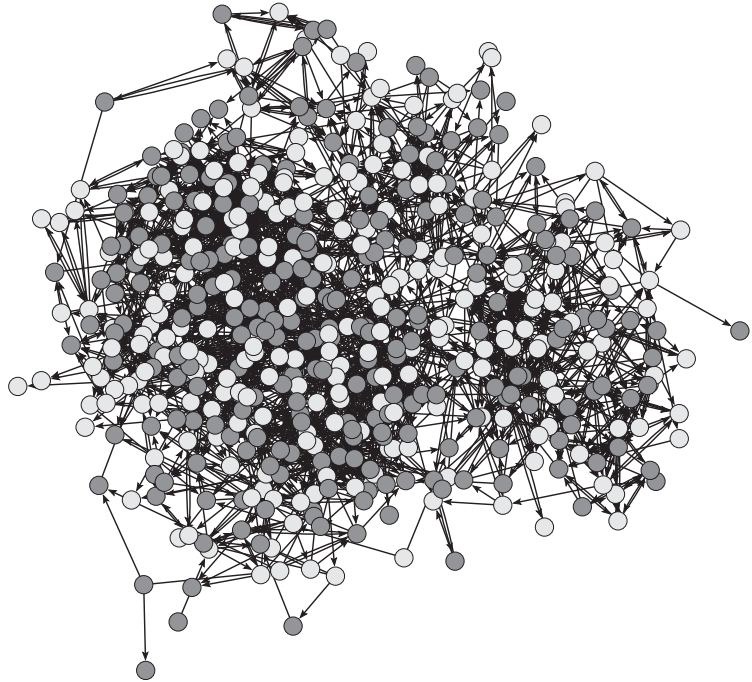}
\end{center}
\caption{\protect
    Illustration of a network\index{network} of different social 
    friendships in 
    a U.S.~School~\cite{addhealth}. Different gray tonalities 
    distinguish between male and female students. Arrows indicate 
    who points who as a friend.}
\label{fig1}
\end{figure}

The use of the network approach to solve problems goes back to the
eighteenth century, when Leonhard Euler wanted to solve the 
``Seven Bridges of K\"onigsberg''\index{Koenigsberg problem} 
problem. The city of K\"onigsberg -- now called Kaliningrad, 
located in Russia -- is set on both sides of Pregel River, having
two islands in between. Both islands are connected with each other
and with both sides of the river by seven bridges,
and Euler posted the problem of proving whether it is possible
to cross each bridge exactly once and return to the starting
point. By solving it, showing that such a walk does not exist, Euler
introduced the concept of {\it graph} as the mathematical object 
composed by nodes (the elements) and edges (the connections) and
founded the so-called graph theory\cite{livro}.
Recently, physicists adopted instead the term 
{\it network} and, with the help of computers, developed 
the network approach to solve a range of new problems spanning 
physics, biology, sociology, and economics.
Since the underlying networks have a very complicated structure as the
one sketched in Fig.~\ref{fig1}, it is common to refer them as complex 
networks.
Several reviews and books on complex network research have been
published~\cite{livro2,albert02,dorogovtsevrev,newman03,boccaletti06,%
dorogovtsev07} in the last decade.

A good way to understand why such an approach is indeed helpful is to
present some of its main achievements.
In this Chapter we will briefly describe the main theoretical and 
numerical tools as well as computational algorithms developed to 
construct and study networks. 
In Sec.~\ref{sec:classification} the main types of networks
are described and in Sec.~\ref{sec:characterization} the
fundamental properties to characterize a network are given.
In Sec.~\ref{sec:families} three families of networks are
described, namely random networks, small-world networks and
scale-free networks and a brief overview over the recent trends
in network approach are given in Sec.~\ref{sec:discussions}.

\section{Classification of networks}
\label{sec:classification}

\begin{figure}[t]
\begin{center}
\includegraphics*[width=12.0cm,angle=0]{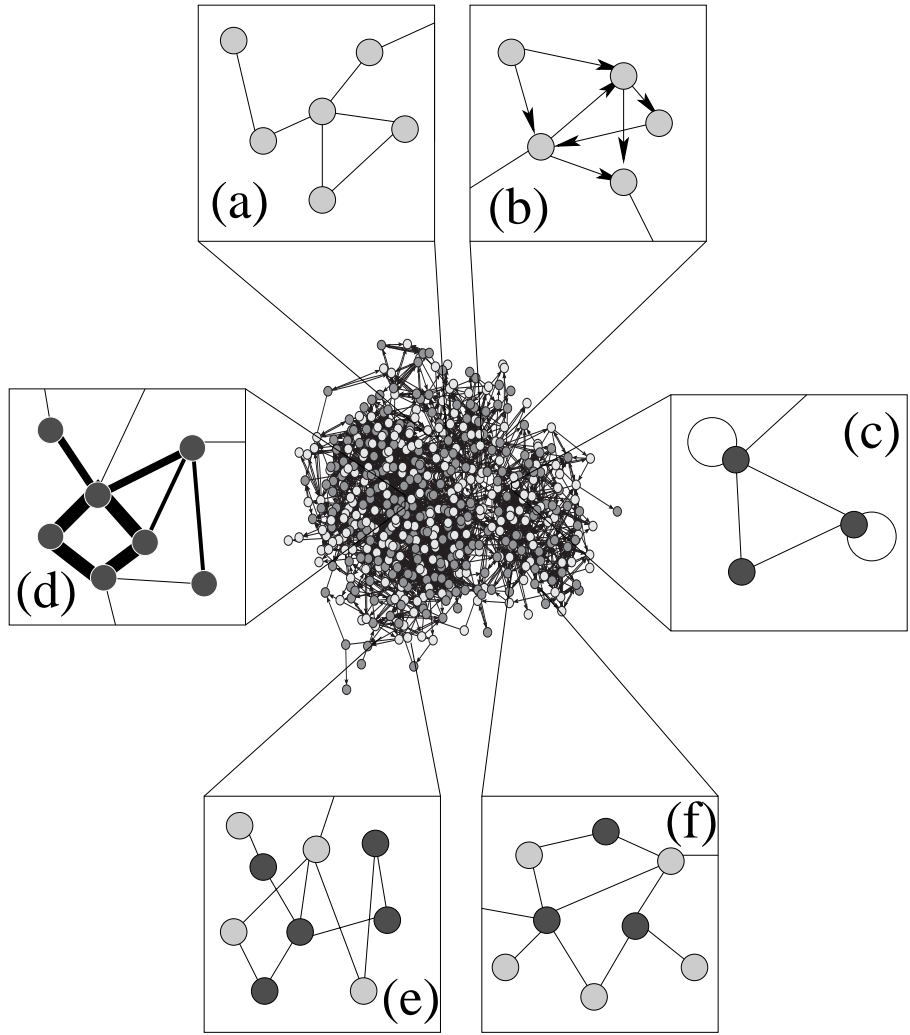}
\end{center}
\caption{\protect
    Classification of networks. Illustration of parts of a network
    that are
    {\bf (a)} undirected,
    {\bf (b)} directed,
    {\bf (c)} with self-connections,
    {\bf (d)} weighted,
    {\bf (e)} monopartite, where the nature of the node does not influence
              the existence of connection and
    {\bf (f)} bipartite, where nodes of one kind only connect to nodes
              of the other kind.} 
\label{fig2}
\end{figure}

A network consists of nodes and edges.
The features we address to the nodes and to the 
edges depend on what we want to study. 
Therefore, the first thing to do when constructing a 
network is to know what kind of components and interactions
between them are we talking about.
Figure \ref{fig2} summarizes the main kinds of nodes and connections
in networks.
\begin{figure}[t]
\begin{center}
\includegraphics*[width=6.0cm,angle=0]{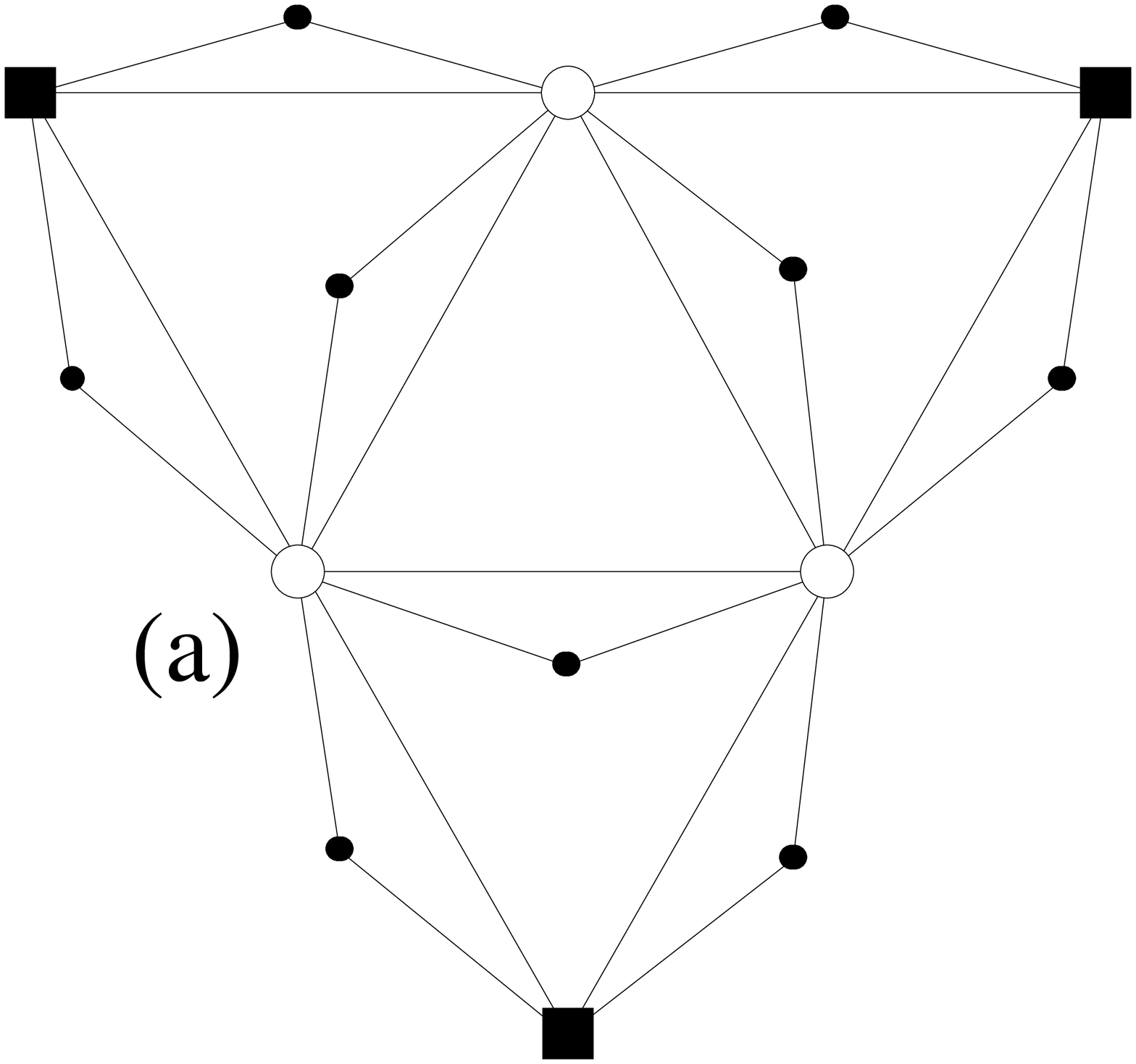}%
\includegraphics*[width=6.0cm,angle=0]{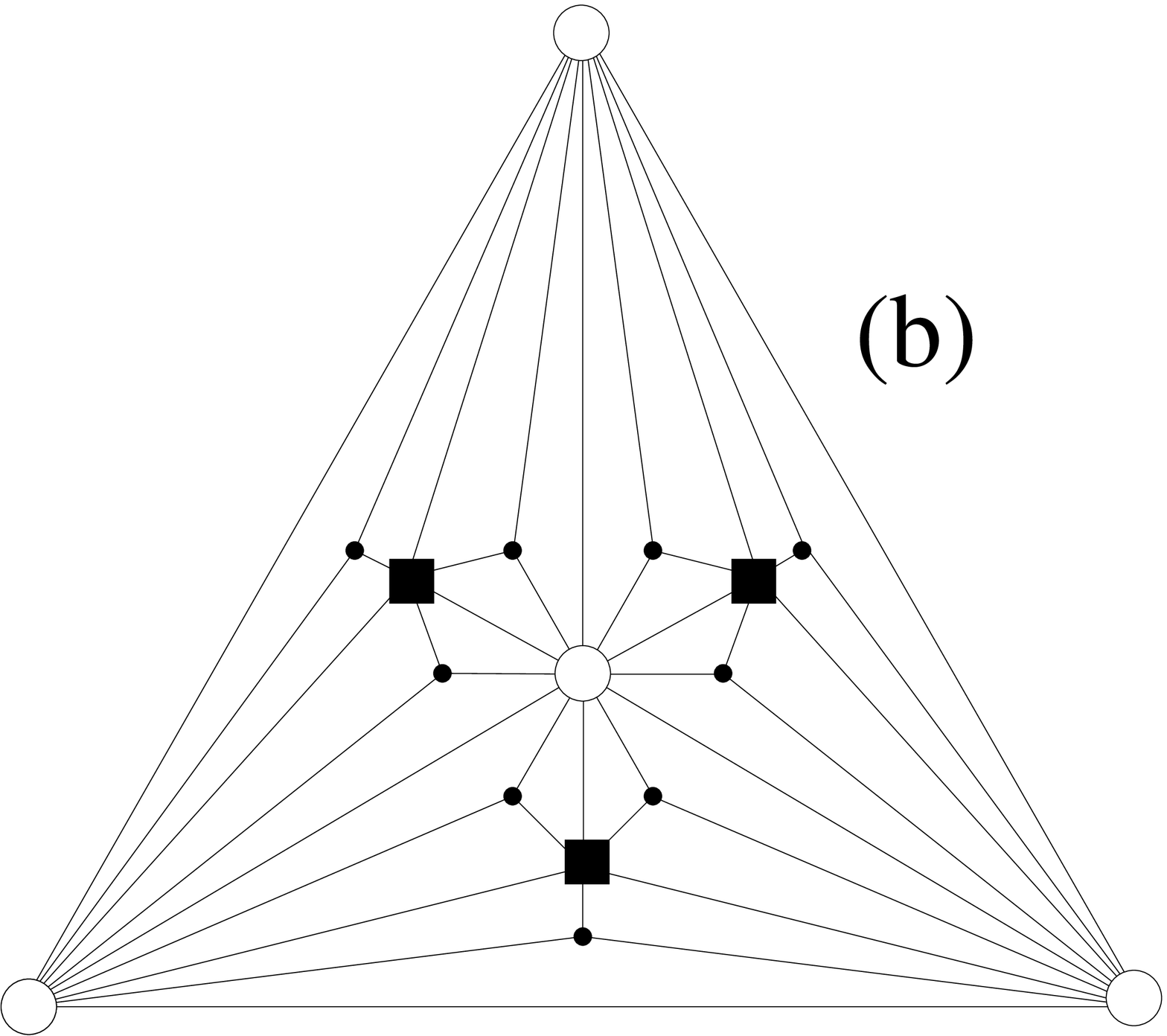}
\end{center}
\caption{\protect
    Two examples of deterministic networks:
    {\bf (a)} the pseudo-fractal network~\cite{dorogovtsevrev} 
              \index{pseudo-fractal network}\index{network, pseudo-fractal}
              and
    {\bf (b)} the Apollonian network~\cite{hans05}.
              \index{Apollonian network}\index{network, Apollonian}
    Different symbols for the nodes represent different generations 
    during the construction procedure (see text).}
\label{fig3}
\end{figure}

We start by discussing the interactions. As in physics, social and
economical systems involve interactions that may or may not be 
symmetric.
For instance, a friendship connection between two persons, $i$ and $j$,
is symmetric, in the sense that there is no preferred direction connecting
$i$ to $j$ or vice-versa. Therefore, edges as the ones illustrated in 
Fig.~\ref{fig2}a are sufficient to represent the connections between
people that are friend of each other. Conversely, if we want to see 
which people a given person {\it knows} inside a social system, then 
there will be for sure some people more famous than other, independent
of their number of friends. Person $i$ knowing person $j$ does not 
guarantee that person $j$ also knows person $i$. A direction must be given and 
therefore one uses arrows to represent interactions, as shown in
Fig.~\ref{fig2}b.

Networks composed by symmetric interactions (edges)
are called undirected 
networks\index{undirected network}\index{network, undirected}, 
while networks composed by arrows
are called directed networks\index{directed network}\index{network, directed}.
While this difference between symmetric and asymmetric interactions
may seem subtle or spurious it has enormous implications in the
way dynamical processes take place on the corresponding networks.
For instance, in rumor propagation or flux of money
or information~\cite{boccaletti06,dorogovtsev07}. 

The two counter parts of symmetric and asymmetric interactions 
appear only in connections joining different nodes. But it may be
the case that self-connections\index{self-connection}, as the ones 
illustrated in 
Fig.~\ref{fig2}c, play an important role in a network. 
Recently it was shown~\cite{hellstein07} that
in a network of authors where connections represent citations
to other authors, self-citations are an important feature to track
the shifting of scientists from one field to another.

Further, after establishing the nature of the interactions,
there is still the possibility to address a value (or weight) to the
connection. 
For instance, the number of phone calls between two persons in a
social system may be regarded as a measure of the friendship
closeness of those two individuals~\cite{martaNature}.
Figure \ref{fig2}d illustrates weighted connections, opposite
to Fig.~\ref{fig2}a where connections are 
unweighted\index{weighted network}\index{network, weighted}.

As for the nature of the nodes they may determine if connections
may exist between them or not. If nodes are people and connections
join the people that are co-authors of the same paper, then no restriction
from the nodes itself is imposed. Some of the persons may be
men and other women. It will make no difference for the
connections, as illustrated in Fig.~\ref{fig2}e. But it may be the case
that the nature of the nodes strongly biases the occurrence of
a connection of some sort between them. For instance, in a
network where nodes are either men and women and connections
represent intimate relations between them, during their life times, 
one expects a stronger tendency for men being attached to women
and vice-versa, as illustrated in Fig.~\ref{fig2}f.
This particular kind of structure where nodes of one sort are connected
to nodes of the other sort is known as a bipartite 
structure\index{bipartite network}\index{network, bipartite}, 
opposite of monopartite 
structures\index{monopartite network}\index{network, monopartite}. 
In general, multipartite networks
are also sometimes addressed, for instance in an ecological system
where connections represent trophic relations between several different
species.

Of course, these features described above may appear solely or combined.
For instance, a network where the interactions measure the number
of phone calls done by a given person $i$ to another person $j$ 
has connections that are directed ($i$ calls $j$) and weighted
(number of phone calls).

Altogether, these network ingredients compose the bulk of the fundamental 
ways for constructing a network underlying a real system. However, as
mentioned earlier they have a very complicated structure, which in
general, is very difficult to be studied analytically. The main reason
lies in the fact that the distribution of nodes and connections is
associated with some stochastic or probabilistic events. 
Therefore, analytical study of networks is usually done in some 
prototypical deterministic networks that contain the structural properties
of the real networks we want to study. 

Deterministic networks are constructed iteratively
by introducing new nodes with a certain deterministic rule.
In Fig.~\ref{fig3} we show two examples of deterministic networks.
The first one is the so-called pseudo-fractal 
network~\cite{dorogovtsev02}\index{pseudo-fractal network}\index{network, pseudo-fractal}: one starts from
three interconnected nodes, and at each iteration
each edge generates a new node, attached to its two vertices.
At a certain iteration $n$, the number of nodes is given
by $N_n=\tfrac{3}{2}(3^n+1)$ and the number of connections by
$V_n=3^{n+1}$.
The second one is the Apollonian 
network\index{Apollonian network}\index{network, Apollonian} which is 
constructed in a different way:
one starts with three interconnected nodes, defining a triangle;
at $n=0$ one puts a new node at the center of the triangle and joins
it to the three other nodes, thus defining three new smaller triangles;
at iteration $n=1$ one adds at the center of each of these three triangles
a new node, connected to the three vertices of the triangle, defining
nine new triangles and so on. In this case one has
$N_n=\tfrac{1}{2}(3^{n+1}+5)$ and $V_n=\tfrac{3}{2}(3^{n+1}+1)$.

In Section \ref{sec:families} we will see that some of the characteristics
of these two networks are ubiquitous in real systems. Next, we will 
introduce the main tools that enable to uncover the structure of
complex networks.

\section{Characterization of networks}
\label{sec:characterization}

After knowing the main ingredients to construct a
network the next task is to know what measurements are needed
to understand the structure. 
In this Section we will briefly address the main
statistical and topological properties in network research. 
For more details the reader may be interested in a
recent review~\cite{boccaletti06}.

The core of all the panoply of properties described below is the so-called
adjacency matrix\index{adjacency matrix}\index{network, adjacency matrix} 
${\cal A}$. It is defined by elements
$a_{ij}$ that are different from zero only if there is one connection
linking $i$ to $j$. If the network is undirected, ${\cal A}$ is 
symmetric ($a_{ij}=a_{ji}$). Further, if the network is unweighted
then the elements are either $0$ (not connected) or $1$ (connected),
while for weighted networks $a_{ij}$ takes real values within a 
given range.

Since networks can be considerably large and adjacency matrices are
typically sparse, when implementing it in a computer program
one usually uses the mapping of the adjacency matrix into
a $2$-linked lisk, which is a $M\times 2$ matrix, {\tt mat\_adj(m,n)}, 
with {\tt m}$=1,\dots, M$ labelling the connections and 
{\tt n}$=1,2$ indicating the two vertices of one connection. 
The connection {\tt m} from $i$ to $j$ is then defined by the two entries 
{\tt mat\_adj(m,1)}$=i$ and {\tt mat\_adj(m,2)}$=j$.
Notice that for undirected networks one needs in fact a $2M\times 2$ 
matrix, since each connection must be defined from $i$ to $j$ and
simultaneously from $j$ to $i$.

In the following we will focus on the simplest case of undirected and
unweighted networks. Extensions to all other types of networks are
straightforward.

\subsection{Degree distribution and correlations}

Within a network, each node $i$ has a certain number of connections which
can be measured by its degree
\begin{equation}
k_i = \sum_j a_{ij} .
\label{degree}
\end{equation}
The degree of a node takes integer values for unweighted networks and 
real values for weighted networks. For directed networks the degree 
in Eq.~(\ref{degree}) equals the so-called outgoing degree, $k_i^{(out)}$. 
The number of incoming connections is then measured by 
$k_i^{(in)}=\sum_j a_{ji}$.

With the degree of each node one easily computes the degree 
distribution\index{degree distribution}\index{network, degree distribution}
$P(k)$, which equals the fraction of nodes having degree $k$. 
As we will see this topological quantity is able to distinguish between
different network families. In some cases, it may yield large fluctuations, 
especially when networks are not large enough as frequently happens in
empirical systems. For that, one may prefer to compute the cumulative
distribution $P_{cum}(k)=\sum_{k^{\prime}>k} P(k^{\prime})$.

The degree distribution, however, does not tell about the correlations
between the nodes. For instance, what is the probability that a node 
with degree $k$ is connected to a node with degree $k^{\prime}$? 
The answer of course, is the conditional probability 
$P(k^{\prime}\vert k)$, that also defines the average
degree of the nearest neighbors of nodes with degree $k$
\begin{equation}
k_{nn}(k) = \sum_{k^{\prime}} k^{\prime} P(k^{\prime}\vert k).
\label{correlation}
\end{equation}

If there are no correlations, then the conditional probability is independent
of $k$ and $k_{nn}(k)=\langle k^2\rangle / \langle k\rangle$.
If the network is correlated, then $k_{nn}(k)$ varies with $k$ and the two
cases may be distinguished. When $k_{nn}(k)$ increases with $k$,
i.e.~nodes with similar degrees tend to be connected, the network is called
assortative. This is the most typical situation found in social networks. 
Oppositely, if the less connected nodes are preferentially attached
to the most connected ones, $k_{nn}(k)$ decreases with $k$ and the
network is called disassortative.
Biological networks are examples of disassortative networks.

\subsection{Clustering coefficient and cycles}

From the degree distribution and degree-degree correlations we are
able to know how many neighbors we should expect to observe when
picking randomly a node in the network and what is the expected
degree of its neighbors.
The next question is: {\it which} nodes are neighbors of the neighbors
of a given node? Are they also neighbors of the given node?

To answer these questions, other quantities are introduced.
To know which neighbors are neighbors between them, one just measures 
the fraction of connection among the neighbors of a given node from 
all possible connections, yielding the so-called
clustering coefficient~\cite{watts98}\index{clustering coefficient}%
\index{network, clustering coefficient}.
A node $i$ with $k_i$ neighbors yields $k_i(k_i-1)/2$ possible connections 
between them in the case of an undirected network, $k_i(k_i-1)$ for a 
directed one, and therefore, if there are only $m_i$ connections between 
the neighbors the (undirected) clustering coefficient is just
\begin{equation}
C_3 = \frac{2m_i}{k_i(k_i-1)} = 
      \frac{\sum_{j,n}a_{ij}a_{jn}a_{ni}}{k_i(k_i-1)} ,
\label{cluscoeff} 
\end{equation} 
where the last member shows explicitly how to count the number
of existing edges between neighbors of node $i$.

The quantity $m_i$ in Eq.~(\ref{cluscoeff}) counts
the number of cycles with size three (triangles) containing
node $i$.
That's the reason why we label the clustering coefficient with the 
subscript $3$.
Triangles are the smallest cycles within a network.
But one could think of cycles with larger size to know about non-local
features of the network.
In bipartite networks for instance such larger cycles are important, 
since they have no triangles and therefore the clustering coefficient
in Eq.~(\ref{cluscoeff}) cannot provide any useful 
information\cite{lind05}.

In general, the number of cycles of size $n$ containing a specific
node $i$ are given by
\begin{equation}
N_c(i,n) = \sum_{j_1\dots j_{n-1}} a_{ij_1}a_{j_1j_2}\dots a_{j_{n-1}i}.
\label{gencluscoeff}
\end{equation}
In practice it is very expensive to count $N_c$ for arbitrary
$n$. Two different approaches are used instead. 
One is to use Monte Carlo algorithms\cite{rozenfeld05}, 
based on large samples of random walks trough the connections and counting 
how many times one returns to the starting node. The other is to 
estimate the number of larger cycles by counting only small cycles, 
typically triangles and squares\cite{lind05,vazquez05}.

\subsection{Average shortest path length}

The cycles mentioned previously are closed paths whose statistical
features provide information about the underlying network beyond 
the node's local vicinity.
Related with cycles is the question of how far are two given
nodes in a network, which leads naturally to the concept of average 
shortest path length.

The average shortest path length $\ell$ is the average number of 
connections joining two randomly chosen nodes.
Of course, $\ell$ increases with the network size $N$ and 
an important point is to ascertain how fast is this increase.
If $\ell$ increases slowly, typically with $\ln{N}$, then 
two given nodes are typically near each other, illustrating what is
called the small-world effect~\cite{watts98}.

Though simple to understand, the average shortest path length
is not trivial to compute efficiently, since it concerns the 
determination of the shortest path\index{shortest path}%
\index{network, shortest path} from all existing paths
between two nodes.
An efficient algorithm to compute $\ell$ is the 
Dijkstra algorithm\index{Dijkstra algorithm}
which is a sort of a burning algorithm of the breadth-first search 
type.
One starts from a given node and visits only one single time
each one of the other nodes in the network, first the nearest 
neighbors ($\ell=1$), then the next-nearest neighbors ($\ell=2$)
and so on.
An implementation in Fortran of the Dijkstra algorithm is as
follows:
\begin{verbatim}
     avpl = 0.0d0
     do 1000 node=1,Nnodes   !Loop in the nodes
        avplnode    = 0.0d0
        flg         = 0
        pathlength  = 0
        path        = 0
        nactualsite = 1
        site(1)     = node
        do i=1,Nnodes
           vs(i) = 1
        enddo
        vs(site(1)) = 0
        do while (flg.eq.0)
           pathlength = pathlength + 1
           nsite      = 0
           do actualsite=1,nactualsite !Loop on node layer 
              do 1100 i=1,Nedges    !Loop in the connections
                 if(mat_adj(i,1).ne.site(actualsite))goto 1100
                 if(vs(mat_adj(i,2)).eq.0)goto 1100
                 path             = path  + 1
                 avplnode         = avplnode+float(pathlength)
                 nsite            = nsite + 1
                 nextsite(nsite)  = mat_adj(i,2)
                 vs(mat_adj(i,2)) = 0
1100          continue
           enddo
           if(nsite.eq.0)then   !All nodes where visited
              flg=1 
           else                 !... if not ...
              nactualsite = nsite
              do i=1,nsite
                 site(i)=nextsite(i)
              enddo
           endif
        enddo
        avpl=avpl+avplnode/float(path)
1000 continue     !END loop in the nodes
     avpl = avpl/float(Nnodes)
\end{verbatim}

Such routine is intended for empirical networks, where the
adjacency matrix is taken as input. For artificial networks,
the algorithm could be implemented more efficiently
by constructing a distance matrix, while saving the entries
for the adjacency matrix during the construction procedure.

\subsection{Community decomposition}

The clustering coefficient mentioned above can be formulated
in a more general form.
How large can a strongly connected vicinity of a given node be?
This question leads to the concept of community:
some subset of nodes in the network whose density of connections 
between them is larger than the density of connections to or from 
the exterior of that subset. 
\begin{figure}[t]
\begin{center}
\includegraphics*[width=11.5cm,angle=0]{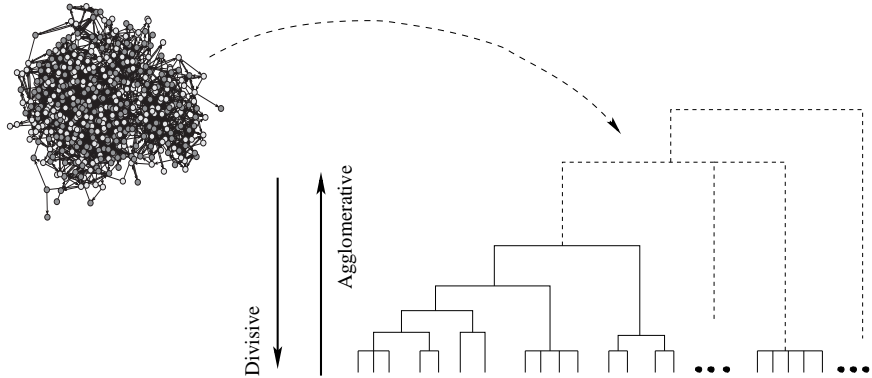}
\end{center}
\caption{\protect
    Sketch of the mapping of a network (left) into a dendrogram
    (right). The dendrogram can be constructed using either agglomerative
    or divisive algorithms (see text).}
\label{fig4}
\end{figure}

Having a network with $N$ nodes and $M$ connections one can define
the average density of connections as $\rho_{Net}=2M/(N(N-1))$.
Similarly, any subset of this network with $n$ nodes and $m_{in}$ 
connections between them would have an average density 
$\rho_{in}=2m_{in}/(n(n-1))$. If, besides the $m_{in}$ 
connections between the nodes composing the subset, there are
other $m_{out}$ connections between the $n$ nodes and the remaining
$N-n$ nodes outside the subset, the corresponding density is 
$\rho_{out}=m_{out}/(n(N-n))$.
For the subset to be a community the following condition must
be satisfied\cite{reichardt04}
\begin{equation}
\rho_{out} < \rho_{Net} < \rho_{in}.
\label{conditionComm}
\end{equation}
This condition, however, does not define a community in a unique way.
Different community 
structures\index{community decomposition}%
\index{network, community decomposition}, in size and number, can 
simultaneously satisfy condition (\ref{conditionComm}) for the
same network.
Thus, different methods to detect communities have been 
proposed~\cite{livroSoc,girvan02,newman04,palla05}.

The traditional method is the so-called hierarchical 
clustering\cite{livroSoc} that maps the network into a 
dendrogram\index{dendrogram}, as illustrated in Fig.~\ref{fig4}.
One starts with all nodes of the network and no edges (bottom of
the dendrogram), addressing to each pair of nodes, $i$ and $j$,
a weight $w_{ij}$ that measures how close connected are the nodes.
This weight could be given by the inverse of the shortest
path length joining the two nodes, $w_{ij}=1/\ell_{ij}$.
Then iteratively, one adds the links between pairs of nodes
following the decreasing order of the weights, which leads to
the agglomeration of small groups of nodes into larger and
larger communities. This method is thus agglomerative.

The dendrogram can also be constructed in the reverse direction
(top-down): one starts with the entire network of nodes and edges
and iteratively cuts the edges, thus dividing the network into
smaller and smaller groups~\cite{girvan02}.
This method is called divisive and the central point is which
edges should be cut at each iteration.
A good criterion is based on the computation of
the so-called betweenness of each edge, that counts the number of
shortest path lengths crossing that edge. Having the betweenness
of each node one follows the decreasing
order of betweenness while cutting the edges\cite{girvan02,newman04}.

Both agglomerative and divisive algorithms for community detection
assume that communities do not intersect with each other. 
However, in real systems this is not the common situation.
To overcome this drawback a different algorithm was 
proposed\cite{palla05}, that enables the detection of communities
that overlap each other. In the heart of this algorithm 
is the concept of $k$-clique, a complete subgraph of size $k$,
i.e.~a subgraph with $k$ nodes where each node is connected to
the remaining $k-1$ nodes. Detecting all $k$-cliques in the network,
a $k$-clique community is then defined as the union of all
$k$-cliques that can be reached from each other.

The investigation of how to detect communities is one of the 
most recent topics in network research and has attracted huge
attention, since communities are ubiquitous in real networking systems 
and are crucial for the understanding of their structure and evolution, 

\section{Three important network topologies}
\label{sec:families}

\begin{table}[b]
\centering
\begin{tabular}{cccc}
  \hline\noalign{\smallskip}
                     & \ \ \ \ Random\ \ \ \   
                     & \ \ \ \ \ Small-world\ \ \ \ \  
                     & \ \ \ \ Scale-free\ \ \ \  \\
\noalign{\smallskip}\hline\noalign{\smallskip}
$P(k)$  & $\hbox{e}^{-\bar{k}}\bar{k}^k/k!$       
                 & $\hbox{e}^{-\bar{k}}\bar{k}^k/k!$             
                 & $2m^2/k^3$        \\
\noalign{\smallskip} 
$C_3$  & $\bar{k}/N$   
              & $C_0(1-p)^3$         
              & $\sim N^{-3/4}$   \\
\noalign{\smallskip} 
$\mathbf{\ell}$ &        
          $\ln{N}/\ln{(pN)}$    &
          $\ln{N}$   &
          $\ln{N}/\ln\ln{N}$   \\
\noalign{\smallskip}\hline
\end{tabular}
\caption{Distinguishing different complex topologies with 
  the degree distribution $P(k)$, clustering coefficient $C_3$ and 
  average shortest path length $\ell$.
  Here $N$ is the total number of nodes, $p$ is the probability for
  two nodes to be connected, $\bar{k}$ is the average number of
  connections per node, $C_0$ is the clustering coefficient of the
  regular network from which the small-world network is constructed, and
  $m$ is the number of initial connections of each new node in a
  scale-free network (see text).} 
\label{tab1}
\end{table}

With the main statistical and topological tools described above
it will be now possible to briefly present the main families
of networks that are usually studied.
In this Section we will describe the fundamental properties of
Random, Small-world and Scale-free networks, and explain how to construct 
them.
Table \ref{tab1} summarizes their fundamental properties.
For the interested reader, the reviews cited 
above~\cite{livro2,albert02,dorogovtsevrev,newman03,boccaletti06,%
dorogovtsev07} provide additional information.

\subsection{Random networks}

Random networks\index{random network}\index{network, random} were 
introduced by Erd\"os and R\'enyi in the late
fifties\cite{erdos} to study organizing principles underlying some
real networks.
To construct them, one defines the probability $p\equiv p(N)$, function of
the total number $N$ of nodes, which determines the probability
for a pair of nodes to be connected, and applies it to the $N(N-1)/2$
pairs of nodes.
The connections are typically long-range connections, and no degree-degree
correlations are found.
\begin{figure}[t]
\begin{center}
\includegraphics*[width=12.0cm,angle=0]{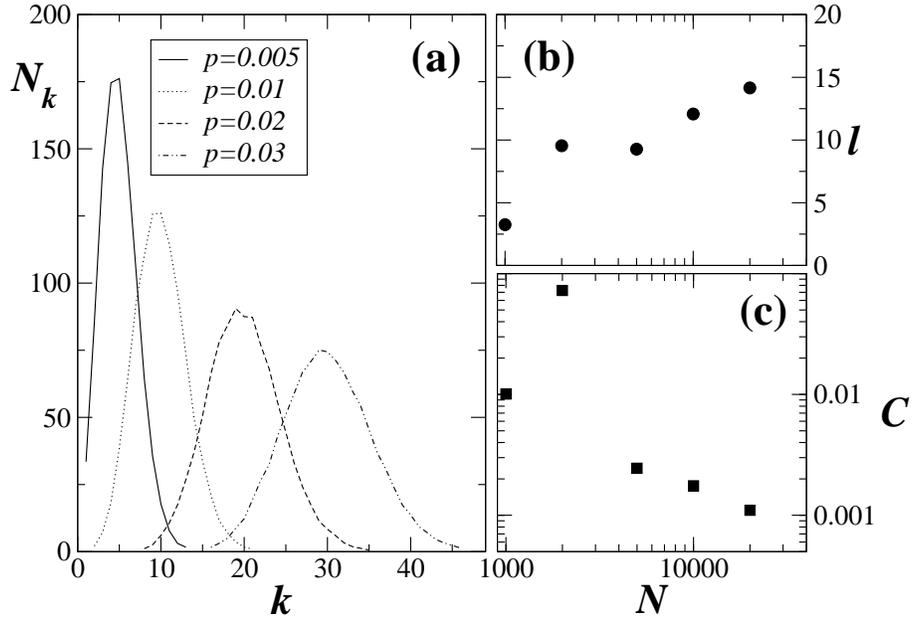}
\end{center}
\caption{\protect
    Characterization of Random Networks.
    {\bf (a)} The degree distribution $P(k)=N_k/N$ for
              different values of the connection probability $p$,
              with $N=2000$ nodes and
    {\bf (b)} the average shortest path length $\ell$ and
    {\bf (c)} the clustering coefficient $C$, as a function of the
              network size $N$ fixing $p=0.01$.}
\label{fig5}
\end{figure}

The degree distribution is typically Poissonian, as shown in 
Fig.~\ref{fig5}a and
an important feature of random networks, which also appears 
in real networks, is their small average shortest path length $\ell$.
As illustrated in Fig.~\ref{fig5}b, typically $\ell$ increases
not faster than logarithmically with the network size.
Another typical feature of Random Networks is their small 
clustering coefficient which typically decreases with the network size 
as $1/N$ for sufficiently large $N$, as sketched in Fig.~\ref{fig5}c.

One main goal in studying random networks is to determine the critical
probability $p_c(N)$, beyond which some specific property is more
likely to be observed, e.g.~the critical probability marking a
transition to percolation~\cite{christensen98}.  

\subsection{Small-world\index{small-world network}\index{network, small-world} 
            networks}

\begin{figure}[t]
\begin{center}
\includegraphics*[width=5.0cm,angle=0]{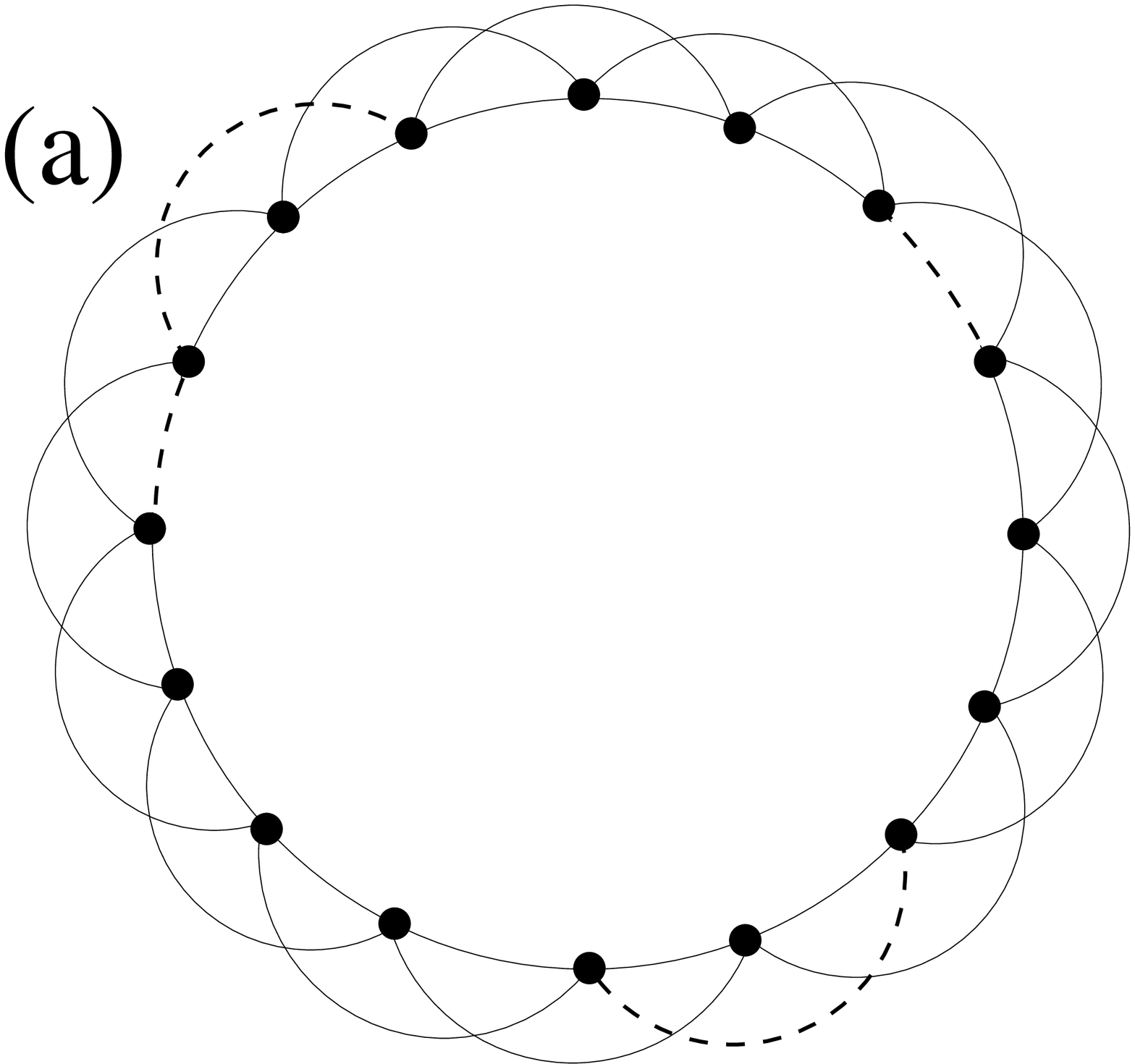}%
\includegraphics*[width=5.0cm,angle=0]{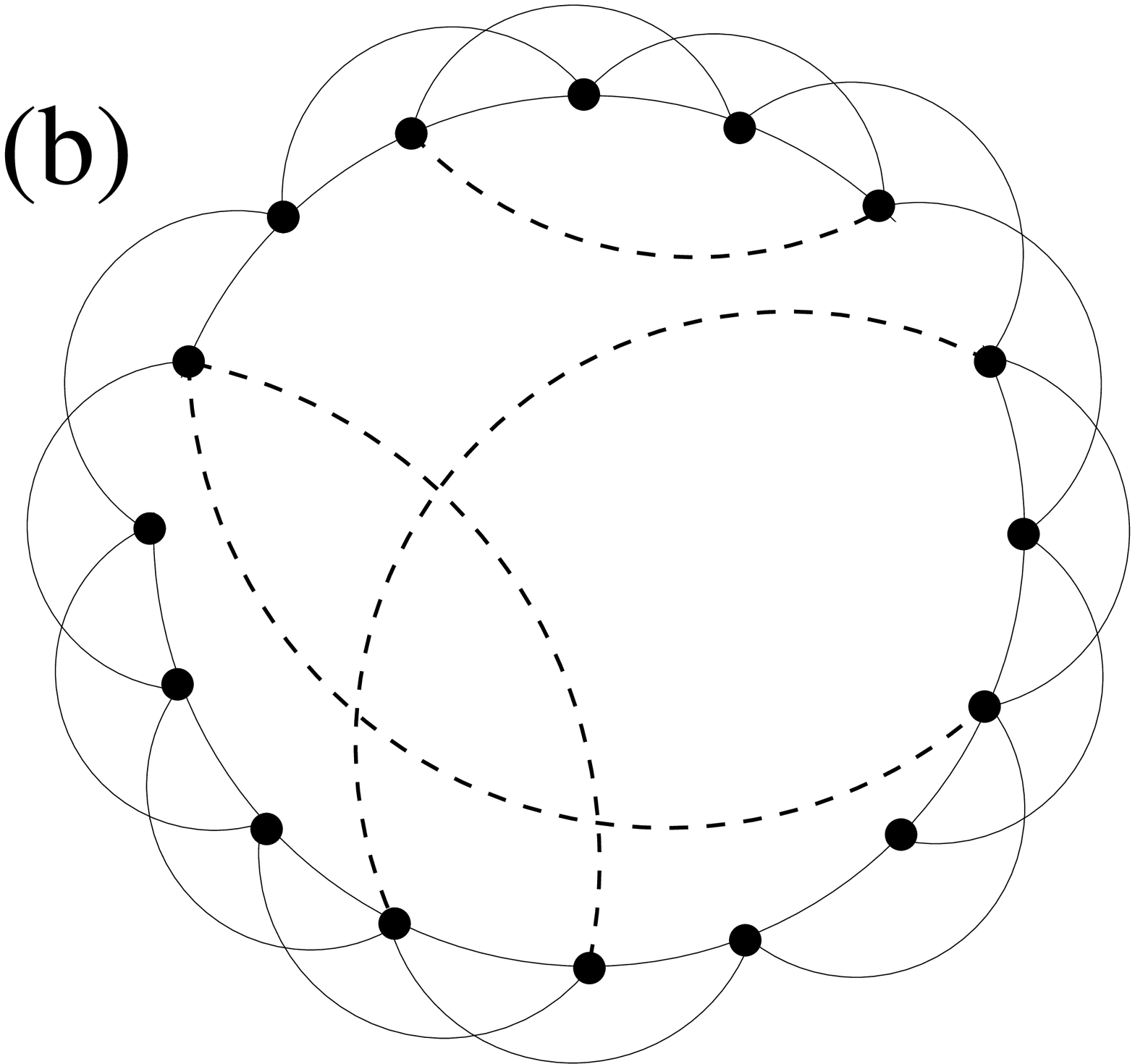}
\includegraphics*[width=12.0cm,angle=0]{fig06c_lind.eps}
\end{center}
\caption{\protect
    To construct a Small-world Network we start with a
    {\bf (a)} regular network and choose a fraction $p$
              of connections to be
    {\bf (b)} rewired randomly to other nodes.
    For a certain range of $p$-values one finds
    {\bf (c)} The Small-world (SW) regime, characterized by
              large clustering coefficients $C$ and small 
              average shortest path length $\ell$, when
              compared to the values $C_0$ and $\ell_0$ found
              for the regular network in (a).
    {\bf (d)} The degree distribution of Small-world networks
              is typically peaked for small $p$ and converges
              to a Poissonian for large $p$.}
\label{fig6}
\end{figure}

While reproducing fairly the shortest path length of many
empirical networks, random networks have also a small clustering
coefficient, which is not typical, e.g., social networks.
In fact, many social networks have simultaneously small $\ell$
and large $C$, i.e.~two persons are typically close to each 
other in the friendship connection web and his/her friends
tend to be also friends of each other.

To reproduce such a topology Watts and Strogatz proposed~\cite{watts98} 
a simple algorithm to construct networks yielding both features.
Starting with $N$ nodes disposed in a chain and connected
with its $k_0$ nearest neighbors (Fig.~\ref{fig6}a) one
{\it rewires} each connection with probability $p$.
As illustrated in Fig.~\ref{fig6}b, such procedure yields a
network composed mainly by short-range connections with
a few number of long-range connections.

While the small number of long-range connections are sufficient to
guarantee a small average shortest path length, the short-range
connections keep the clustering coefficient significantly high,
if one compares with the random network counterpart.
Figure \ref{fig6}c shows both $C$ and $\ell$ as function of the
rewiring probability $p$. For large $p$ the Watts-Strogatz 
procedure yields small $C$ and $\ell$ as in random networks, while
for small $p$ one has large $C$ and $\ell$ as observed in regular
networks. In the middle, typically for 
$0.01\lesssim p\lesssim 0.1$ we find the Small-world (SW) regime
described above.

While sweeping through the spectrum of $p$ values, the degree
distribution varies from a $\delta$-function (regular network)
to an exponential distribution ($p=1$). In the SW regime
the degree distribution is approximately Poissonian, as illustrated
in Fig.~\ref{fig6}d.

Instead of rewiring short-range connections into long-range ones,
an alternate procedure~\cite{newman99} would be to add new 
long-range connections for each existing connection in the network, 
with a probability $p$.
This procedure is more appropriate for most purposes, since
it avoids the possibility of generating disconnected
sets of nodes.  

\subsection{Scale-free networks\index{scale-free network}%
            \index{network, scale-free}}

Both random and small-world topologies have other drawbacks.
First, they do not evolve, the total number of nodes being fixed from
the beginning. Second, there are no criteria from choosing specific
pairs of nodes to be linked. 

In some real networks, that are growing in time, the new nodes tend
to prefer connections with the most connected nodes - so-called hubs -
of the existing network. This happens for instance in
the World Wide Web. New Internet pages tend to link to the most connected
ones.
\begin{figure}[t]
\begin{center}
\includegraphics*[width=12.0cm,angle=0]{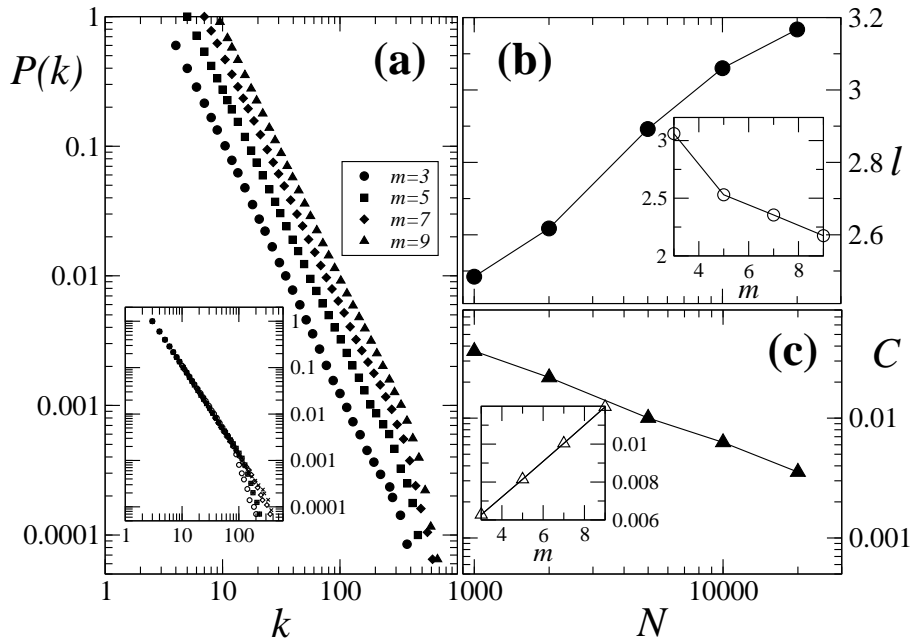}
\end{center}
\caption{\protect
    Characterizing evolving Scale-free networks.
    {\bf (a)} Power-law degree cumulative distribution $P(k)$ for
              different values of the initial number $m$ of 
              connections
              with $N=10^4$ nodes and in the inset for different
              network sizes $N=10^3,5\times 10^3, 10^4$ and 
              $2\times 10^4$ with $m=3$.
    For the same cases one also plots 
    {\bf (b)} the average shortest path length and
    {\bf (c)} the clustering coefficient.}
\label{fig7}
\end{figure}

These two additional ingredients, growth and preferential
attachment, were addressed by Barab\'asi and Albert
\cite{barabasi99} to construct networks with similar features as
the ones of the WWW.
The most crucial feature of these networks is the non-existence of
a characteristic number of neighbours: the degree distribution is a
power-law $P(k)\sim k^{-\gamma}$. Therefore, it is usual to call
such networks, Scale-free networks.

To construct a scale-free network, one starts with a small amount 
of nodes totally interconnected, and adds iteratively one node 
with $m$ connections to the previous nodes, chosen from a probability 
function proportional to their number of connections. 
With this construction one obtains a degree distribution 
$P(k)\propto k^{-\gamma}$, where $\gamma\to 3$ as $N\to \infty$, 
independent of the initial number of fully interconnected nodes and 
$m$, as illustrated in Fig.~\ref{fig7}a.

From Figs.~\ref{fig7}b and \ref{fig7}c one also sees that scale-free 
networks have typically small $\ell$ and $C$, similar to random
networks (see also Tab.~\ref{tab1}). Further, an increase of the 
initial number $m$ of connections tends to decrease $\ell$ and increase
$C$, as shown in the insets of Figs.~\ref{fig7}b and \ref{fig7}c
respectively.

From the computational point of view the preferential attachment
should be implemented by looking at the existing edges instead
of imposing a probability for choosing a node proportional to
its degree. In fact, a random choice on the set of existing edges
is equivalent to a choice on the set of existing nodes proportional
to its degree, and it is much more efficient.

Finally, it is also possible to generate scale-free 
networks, by either imposing {\it a priori} a power-law distribution 
of all connections randomly distributed, or by following a deterministic 
iterative rule for new nodes. 
The first procedure generates what is usually called a generalized
random graph~\cite{albert02}, while the latter concerns among other
the deterministic scale-free networks sketched in Fig.~\ref{fig3}.

\subsection{The landscape analog of complex networks}

Some of the topological and statistical quantities described
in Sec.~\ref{sec:characterization} were not considered
in this Section, namely the degree-degree correlations and 
the community structure. The main reason being that
such quantities may appear differently in any of the above
network topologies.
To end this Section, we present a pictorial overview recently 
proposed~\cite{axelsen06}, to improve the perception 
of all topologies discussed in this chapter.

Lets imagine that the nodes of a certain network are placed over
a landscape. The altitude where each node lies is proportional
to its degree, and neighbours are placed closer than other nodes.
How would the landscape look like for each different topology?

A regular network as the one
picked in Fig.~\ref{fig6}a, where all nodes have the same degree,
would yield a perfect plateau. Other networks with some typical degree,
like random or small-world networks, would have a smooth landscape
with some hills alternating with valleys.
High mountains would appear, of course, in scale-free networks.

The number of hills or mountains would then depend on the
degree-degree correlations. For positively correlated (assortative) 
networks, where nodes of similar degree tend to connect
with each other, a single bump is observed. A classical
example of such a network is the Internet. If, on the contrary,
high connected nodes have low connected nodes as neighbours
(negative correlations), several hills or mountains appear.
Examples of such rough landscape can be found in biological
systems.

Besides, this nice analog not only catches the main features of
the different topologies described above, but also provides new
insight for uncovering the hierarchical structure of the complex 
network~\cite{axelsen06}.

\section{Recent trends in network research}
\label{sec:discussions}

The above Sections hopefully provide a glimpse of
what network approach is about.
What to do next?
What are the main research activities in the field of network
research?
In a broad sense, three fundamental open questions remain in network 
research.
First, how to model empirical networks from fundamental
principles? Second, what new topological or statistical tools
could be introduced to improve the uncovering of network
structure? And third, how does network structure influence
dynamical processes occurring on them?

In the scope of modeling, there has been a huge amount of data
collected from empirical networks
that promoted an accurate comparison between models and 
reality~\cite{newman03}. 
It has been shown that social networks, for instance, have three 
fundamental common features:
they present the small-world effect, have positive correlations
and invariably present a community structure.
Although there are arguments pointing out that all these features could be
consequence from one another\cite{newman03}, 
the modeling of specific social networks reproducing quantitatively all 
these features at once has not been successful.
Recently, however, it was shown that using a new network construction
procedure, based on a system of mobile agents, it is possible to reproduce 
all these features~\cite{gonzalez06}.

Concerning the new topological tools for accessing 
the structure of networks, a review of the most recent achievements
can be found in Ref.~\cite{boccaletti06}.
One such tool was already discussed above and corresponds
to the estimate of cycle distribution in networks.
Recently, a general theoretical picture of a global measure
of increasing order of clustering coefficients according to some
suitable expansion was described~\cite{lind07}, but a close
form for a general clustering coefficient, which should be related
somehow with the community structure, is still to be done.

Finally, the third question on dynamical processes on networks
ranges from rumor or gossip spreading~\cite{lind07b} to 
synchronization phenomena of local oscillators placed at the
nodes of a network~\cite{dorogovtsev07}.
Recently, a simple model of gossip propagation in an empirical
network of friendship connections has shown a new striking
result: there is a non-trivial optimal number of friends that 
minimizes the risk of being gossiped~\cite{lind07,lind07b}.
The study of rumor and gossip propagation is in fact a
particular case of information spreading in networks, that
also includes the study of robust topologies to prevent
epidemic or informatic virus spreading.

All of us are somehow connected.
That is more or less obvious.
But, as we tried to illustrate above, 
understanding more deeply how we are connected will 
probably uncover many features we do not yet know about
us as groups of persons, and give us new ideas to improve
our life in some of the real networks we live in.


\printindex
\end{document}